\documentclass[a4paper, amsfonts, amssymb, amsmath,reprint, showkeys, nofootinbib, twoside]{revtex4-1} 
\usepackage[english]{babel}
\usepackage[utf8]{inputenc}
\usepackage{url}
\usepackage{acro}
\usepackage{soul}
\usepackage[colorinlistoftodos, color=green!40, prependcaption]{todonotes}

\usepackage{amsthm}
\usepackage{mathtools}
\usepackage{physics}
\usepackage{xcolor}
\usepackage{graphicx}
\usepackage[left=23mm,right=13mm,top=35mm,columnsep=15pt]{geometry} 
\usepackage{adjustbox}
\usepackage{placeins}
\usepackage[T1]{fontenc}
\usepackage{lipsum}
\usepackage{csquotes}

\usepackage[pdftex, pdftitle={Article}, pdfauthor={Author}]{hyperref} 
\usepackage{xcolor}

\DeclareAcronym{uv}{
  short = UV,
  long = ultra violet
}

\DeclareAcronym{xas}{
  short=XAS,
  long=X-ray absorption spectroscopy
}

\DeclareAcronym{xses}{
  short=XSES,
  long=X-ray spontaneous emission spectroscopy
}

\DeclareAcronym{atas}{
  short=ATAS,
  long=attosecond transient absorption spectroscopy
}

\DeclareAcronym{trc}{
  short=TRUECARS,
  long=transient redistribution of ultrafast electronic coherence in attosecond Raman signals
}

\DeclareAcronym{ci}{
  short = CI,
  long = conical intersection
}
\DeclareAcronym{pes}{
  short = PES,
  long = potential energy surface
}

\DeclareAcronym{rwa}{
  short=RWA,
  long=rotating wave approximation
}

\DeclareAcronym{fc}{
  short=FC,
  long=Franck-Condon
}

\begin{document}
\title{Capturing Fingerprints of Conical Intersection: Complementary Information of Non-Adiabatic Dynamics from Linear X-ray Probes}
\author{Deependra Jadoun}    
\author{Mahesh Gudem}
\author{Markus Kowalewski}
    \email[E-mail: ]{markus.kowalewski@fysik.su.se}
    \affiliation{Department of Physics, Stockholm University, Albanova University Centre, SE-106 91 Stockholm, Sweden}

\date{\today} 

\begin{abstract}
Many recent experimental ultrafast spectroscopy studies have hinted at non-adiabatic dynamics indicating the existence of conical intersections, but their direct observation remains a challenge.
The rapid change of the energy gap between the electronic states complicated their observation by requiring bandwidths of several electron volts. In this manuscript, we propose to use the combined information of different
X-ray pump-probe techniques to identify the conical intersection.  We theoretically study the conical intersection in pyrrole using transient X-ray absorption, time-resolved X-ray spontaneous emission, and linear off-resonant Raman spectroscopy to gather evidence of the curve crossing.
\end{abstract}


\maketitle

\section{Introduction} \label{sec:introduction}

\Acp{ci} \cite{domcke11ws, baer2006beyond, yarkony98acs}, though once considered rare, govern the outcome of many of the photochemical reactions, according to modern photochemistry \cite{ci_1,ci_2,ci_3}.
They appear in a molecule when two or more \acp{pes} cross and enable the possibility of electronic de-excitation without radiative emission. These non-radiative transitions result in the breakdown of the Born-Oppenheimer approximation. Consequently, the electronic and nuclear degrees of freedom are strongly coupled in the vicinity of the \ac{ci}. It is now well established that the \acp{ci} play a vital role in several photochemical processes such as photosynthesis \cite{phot_synth_1}, primary event of vision \cite{vision}, photochemical formation of DNA lesions \cite{dna_dam}, and photochemistry of individual nucleobases \cite{nuc_bas}. However, the direct experimental detection of \acp{ci} in molecular systems is still challenging. This is mainly because the energy gap between \acp{pes} near the \ac{ci} decreases rapidly. Ultrashort pulses with adequate resolution in both energy and time domain are required to observe them.

With the advent of X-ray lasers, it is possible to probe such fast processes that occur on the timescale of femtoseconds and attoseconds using ultrafast pump-probe spectroscopy \cite{domcke12arpc, sala16cp, wu19pccp, kremar14jpb, breckwoldt20rxiv, duan20sa, smolared10pccp, sun20jcp, galbraith17nc, worner11sj,baekhoj18prl}. It has been recently shown experimentally that \ac{atas} can aid the observation of non-adiabatic processes \cite{kobayashi19sj,timmers19nc, oliver15jpcb,zinchenko21sj}. Time-resolved photo-electron spectroscopy \cite{neville18prl,bennett16jctc}, and X-ray diffraction \cite{bennett18pnas} have also been proposed to study the behaviour of system in the vicinity of a \ac{ci}.

Spectroscopic techniques that involve attosecond pulses provide a high resolution in the time-domain,
and cover a broad range of frequencies, making them a promising tool to measure the rapidly
varying energy gap that comes with \acp{ci}.
However, their experimental availability can be a limiting factor. 

In the following, we show theoretically how a combination of femtosecond based X-ray probe methods may 
be used to detect the non-adiabatic dynamics in a molecule.
Our chosen example molecule, pyrrole, exhibits a \ac{ci} in the
photodissociation pathway of the hydrogen atom. The branching ratio between the electronic states
in the vicinity of the \ac{ci} is on the order of a few percent, making it a rather challenging
example \cite{domcke_dyn}. The goal of the present work is to show that it is possible to locate a \ac{ci} even when there is an unbalanced branching of the electronic state population in its vicinity, using widely used femtosecond X-ray probe methods. A purely diabatic or purely adiabatic population transfer poses a challenge for spectroscopic X-ray techniques to identify
the existence of the second electronic state involved in the \ac{ci}. Three different techniques have been employed to study the photo dynamics of pyrrole involving a \ac{ci} between ground and excited electronic state. The methods used here are as follows, time-resolved \ac{xas}, \ac{xses}  and recently proposed technique, \ac{trc}. We briefly introduce the features of the methods used in this manuscript in the following. 

In the case when there is an unbalanced branching, one needs to look for the indirect signatures of a \ac{ci} to detect its occurrence. We construct the transient \ac{xas} spectrum using femtosecond X-ray probe-pulses to observe the change in transition dipole-moment in the vicinity of a \ac{ci} in pyrrole. Femtosecond pulses are sufficiently narrow in the frequency domain ($\leq 1$\,eV, as compared to attosecond pulses) to selectively probe the transitions which are distributed over a few electron volts. 
The intensity of the \ac{xas} signal depends on the valence-to-core transition dipole moments that the probe pulse is resonant with, as well as the population of the valence states.
It is possible to construct \ac{xas} spectra using attosecond pulses, but it may contain signatures originating from other resonant valence-to-core transitions, further complicating the interpretation of the spectrum.

Time-resolved \ac{xses} \cite{santoro00jcp,patuwo13jcp} does not solely depend upon the branching of the wave packet
and may provide additional information about the curve crossing. 
In the \ac{xses} spectrum, the energy gap in the vicinity of the \ac{ci} may be observed directly independent of the population of the involved valence states and is expected to yield information
about the shape of the \acp{pes} involved.
In time-resolved \ac{xses}, pyrrole is excited to a nitrogen 1s core-hole state, which is followed by the spontaneous emission of a photon. The 1s core-hole states of nitrogen have a lifetime of 7\,fs \cite{inhester19jcp,hargenhahn01jpca,schoffler08sj}, and the spontaneous emission takes place within this time period after excitation.

Probing the creation of electronic coherences in the vicinity of \ac{ci} provides a more direct signature of \acp{ci}. Several X-ray based spectroscopic techniques have been proposed theoretically to observe this phenomena \cite{hua16sdus,cho20jpcl_2,zhang14pccp,bennett16jctc}. One such scheme is \ac{trc} \cite{kowalewski15prl,keefer20pnas,cho20jpcl}, which is based on an off-resonant linear Raman process.
In this scheme, a hybrid probe-pulse sequence is used instead of a single probe pulse to construct the signal where the redistribution of photons maps out the time varying energy gap via the Raman shift.
\Ac{trc} is only sensitive to the coherences and is thus  expected to yield accurate information about the passage through the \ac{ci}. 
However, to obtain the maximum intensity in the signal, a balanced branching at the \ac{ci} is required.
The result from \ac{trc} signal can be considered confirmation  for the other two methods.
The use of hybrid pulses in \ac{trc} and homodyne detection scheme makes it experimentally more challenging, whereas \ac{xas} and \ac{xses} have been used successfully earlier in experiments to study photochemical processes \cite{andel97bio,kochendoerfer96jpc}.

We show that the three methods deliver complementary information that can be used to observe the direct (generation of vibronic coherence and state resolved spectra) and indirect (change in the transition dipole moments) effects of a \ac{ci} between two states in the pyrrole molecule.
The paper is structured as follows: section \ref{sec:model} presents the details of the system used for the signal calculations. In section \ref{sec:signal}, the analytical expressions for spectroscopic signals and Hamiltonian for the various pump-probe methods are discussed. Section \ref{sec:methods} contains the details about the software and programs used to carry out the simulations. In section \ref{sec:result}, spectra calculated using different pump-probe methods are presented along with the discussion. Finally, the main conclusions drawn by comparing the results from used methods are presented in section \ref{sec:conclude}.

\section{Model} \label{sec:model}
\begin{figure}[t]
  \centering
  \includegraphics[width=8cm]{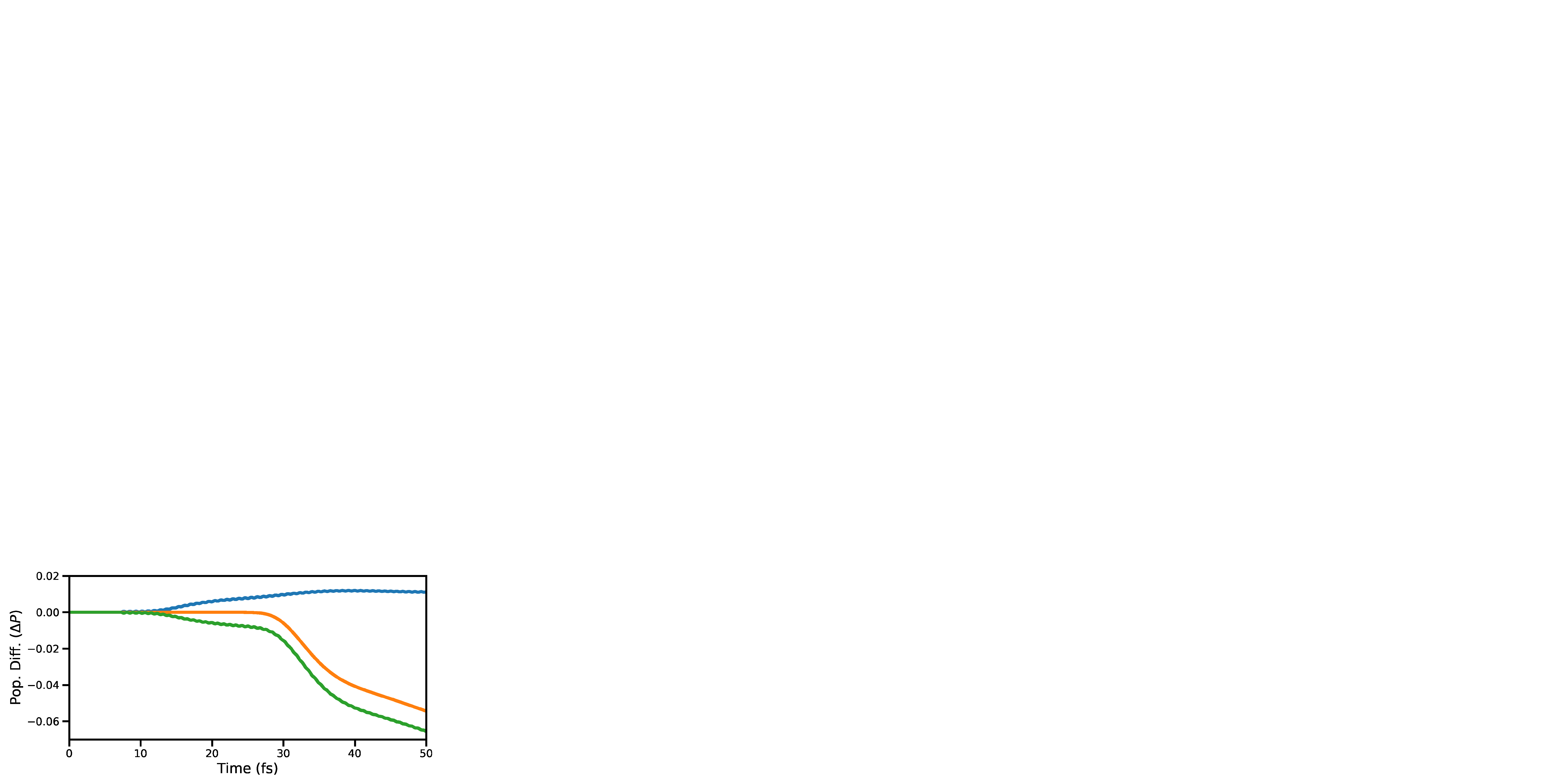}
  \caption{The plot shows the change in population $\Delta P$ after the pump-pulse ends for state $S_0$ (blue curve), state $\pi\sigma^*$ (green curve), and the total population of the system (orange curve). When the wave-packets reach the vicinity of \ac{ci} around 10\,fs, the population branches out, and consequently, the population for the $\pi\sigma^*$ starts to decrease, and the population for $S_0$ starts increasing. The system's total population starts to decay around 30\,fs when the wave-packets hit the boundary of \ac{pes}. The wave-packets are absorbed by the perfect-matched-layer placed at the edge of the \ac{pes}.
} \label{fig:pops}
\end{figure}
\subsection{Photochemistry of pyrrole}
Pyrrole is a nitrogen-containing five-membered heterocyclic aromatic compound, which has been used extensively as a model system for exploring non-adiabatic dynamics \cite{pyr_mod1,pyr_mod2,pyr_mod3,pyr_mod4,pyr_mod5,pyr_mod6,domcke_dyn}. The low-lying bright states of pyrrole emerge around 6 eV \cite{pyr_spec1,pyr_spec2}. It has been challenging to assign the corresponding peaks in \ac{uv} absorption spectrum. This is mainly due to the high density of energy levels lying around the bright states \cite{pyr_spec_chal1,pyr_spec_chal2}.

The \ac{uv}-induced photochemistry of pyrrole involves NH detachment process through the manifold of four lowest singlet excited states, $^1$A$_2$($\pi\sigma^*$), $^1$B$_1$($\pi\sigma^*$), $^1$B$_2$($\pi\pi^*$), and $^1$A$_1$($\pi\pi^*$) \cite{pyr_ex_st,domcke_dyn}. The two $\pi\pi^*$ states have been assigned to the above-mentioned absorption bands around 6 eV. The molecules in these bright states undergo non-radiative decay to the $\pi\sigma^*$ states along the out-of-plane ring deformation coordinate \cite{ring_def1}. The latter states have been found to be repulsive along the NH stretching mode \cite{pyr_mod2,vib_int3,vib_int4}. The ring deformation mechanism was also shown to be feasible on the $\pi\sigma^*$ states \cite{ring_def1}. Along the dissociation process, the ground state energy increases and forms a CI between $S_0$ and $\pi\sigma^*$ states \cite{pyr_mod2}. The nature of the $\sigma^*$ orbital was also shown to be varied during the detachment process. At the \ac{fc} geometry, it is of 3s Rydberg type which changes to valance $\sigma^*$ upon NH stretching and eventually becomes a 1s orbital of the hydrogen atom. This Rydberg-to-valance orbital transformation rationalises the presence of the dissociation barrier on the $\pi\sigma^*$ states \cite{pyr_mod2}. The photodissociation dynamics of pyrrole involving $S_0$/$\pi\sigma^*$($^1$A$_2$ and $^1$B$_1$) CIs have been theoretically investigated by Domcke and co-workers \cite{domcke_dyn}. The study employs the vertical excitation model and simulates the time-dependent dynamics of pyrrole on each $\pi\sigma^*$ state separately. Recently, optical cavities have been found to be hugely influencing the photolysis reaction dynamics in the $\pi\sigma^*$($^1$B$_1$) state \cite{mg_mk}. The present work considers the photodynamics of pyrrole in the $\pi\sigma^*$($^1$B$_1$) state, referred to as just $\pi\sigma^*$ in the rest of the paper, as a model for the study. We have employed a simplified model by considering only the stretching coordinates corresponding to the NH bond. This reduced dimensionality model has been found to be describing the photofragmentation features of pyrrole qualitatively \cite{domcke_dyn}. 

The main objective of the paper is to show how femtosecond based X-ray probe methods can be used to detect the CIs in molecular systems. Therefore, we include only $S_0$ and $\pi\sigma^*$ states, and the vibronic couplings of them with the other valance excited states have been neglected. These additional vibronic interactions \cite{vib_int1,vib_int2,vib_int3,vib_int4} and the other molecular modes like out-of-plane ring deformation \cite{ring_def1}, as mentioned above, may also play a role in describing the dissociation dynamics of pyrrole. Our model does not consider these effects. The treatment of all these couplings in an extended molecular mode space would be necessary for a fully quantitative description of the time-dependent photodissociation dynamics of pyrrole \cite{vib_int1,vib_int2,vib_int3,ring_def1,vib_int4}.

A detailed discussion on the \acp{pes} of valance states employed in the current work can be found in a recent study investigating the cavity-modified dynamics of pyrrole \cite{mg_mk}. We provide the details very briefly here.
Previous theoretical studies have established that the hydrogen elimination reaction dynamics in the $\pi\sigma^*$ state involves hydrogen in-plane and out-of-plane detachment motions \cite{domcke_dyn,mg_mk}. These coordinates, which will be denoted as $R_1$ and $R_2$ refer to the tuning and coupling coordinates, respectively. The computed \acp{pes} along the two reaction coordinates have been used to construct various spectroscopic signals from the quantum dynamics simulations. The valence states are considered to be in the diabatic basis.
When the wave packets reach the \ac{ci}, there is a diabatic population transfer (around $7 \%$ of population in excited state), mediated by the diabatic coupling between the states.
\begin{figure}[t]
  \centering
  \includegraphics[width=8.5cm]{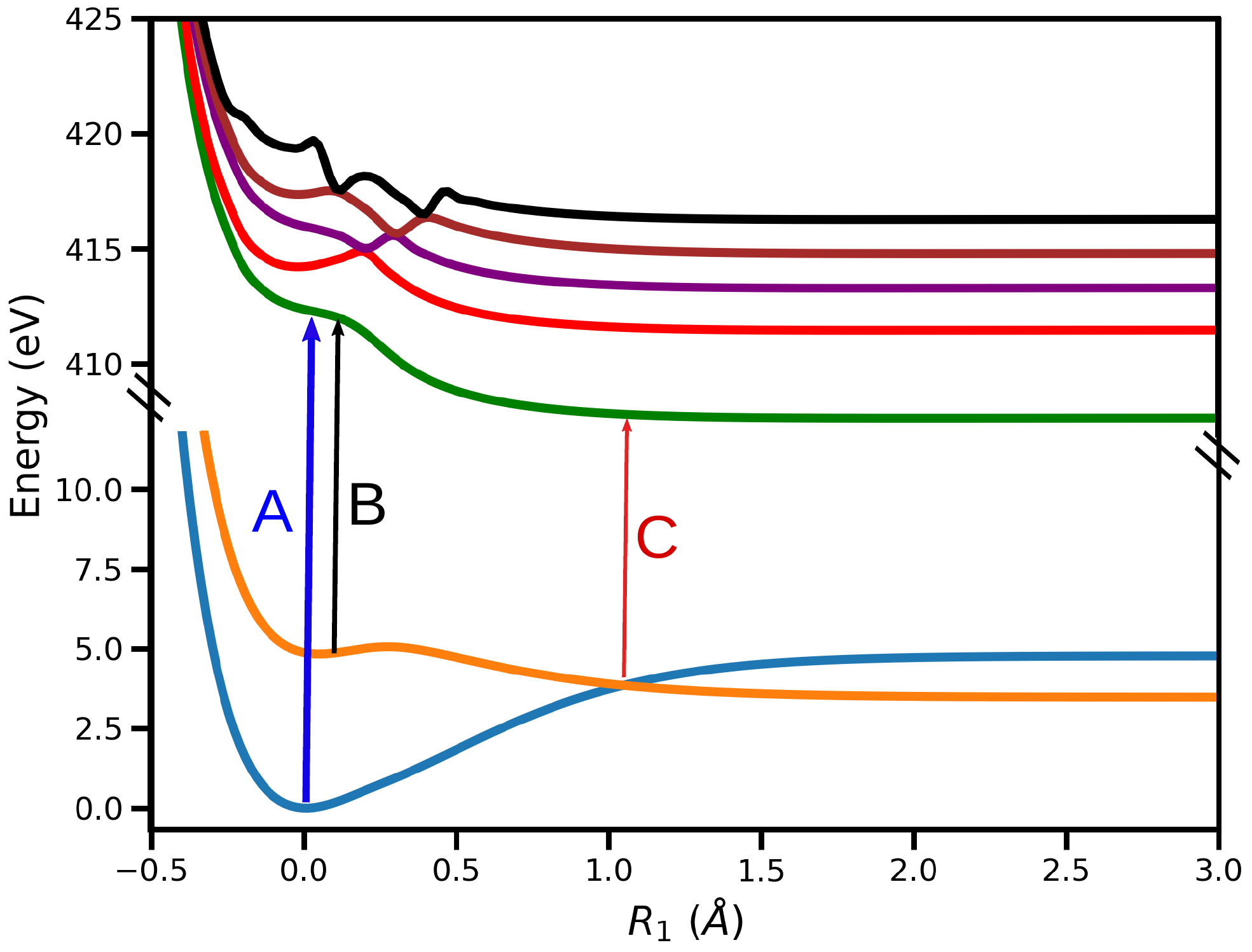}
  \caption{1D cuts of the \acp{pes} of $S_0$ (blue), $\pi\sigma^*$ (orange) and five core-hole states (Nitrogen $1s$ core-hole states) of the pyrrole molecule. These curves show variation of the energy of the states along the reaction coordinate $R_1$. The transitions from the valence-states to the first core-hole state $C_1$ are shown by various arrow. The transition between $S_0$ and $C_1$, represented by A, has approximate energy of 412\,eV. The transition represented by B between $\pi\sigma^*$ and $C_1$ has approximate energy of 407\,eV \cite{shimada14jcp} and transition corresponding to the \ac{ci} region is denoted by C and has approximate energy of 404\,eV.} \label{fig:pes}
\end{figure}
The wave-packets on the $\pi\sigma^*$ state \ac{pes} arrive in the vicinity of \ac{ci} at around $10 \ \mathrm{fs}$ and then starts to branch into $S_0$, which can be seen in the population difference curves shown in Fig. \ref{fig:pops} for $S_0$ state (blue curve) and $\pi\sigma^*$ state (green curve). The population difference for a particular state is calculated by subtracting the maximum population from the population at each time instance after the end of pump-pulse. The total population (orange curve) remains constant until the wave-packets reaches the boundary of the \ac{pes} and gets absorbed by perfectly matched layer \cite{nissen10jcp}. To analyze the behavior of the system near \ac{ci}, we use five core-hole states, which correspond to Nitrogen $1s$ orbital. The 1D strips of the \acp{pes} are shown in Fig.\ \ref{fig:pes}. The two valence states are denoted by $S_0$ and $\pi\sigma^*$ while the lowest core-hole state is denoted by $C_1$.
\subsection{Preparation of the system}
\begin{figure}[t]
  \centering
  \includegraphics[width=4cm]{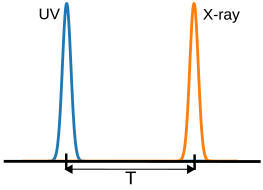}
  \caption{The schematic of the Pump-probe scheme used for the construction of signal. 
  Here, $T$ is the delay time between the pump and the probe pulse.} \label{fig:pul}
\end{figure}

A pump-probe scheme is used in the construction of the signals, as shown in Fig.\ \ref{fig:pul}.
An \ac{uv} pump-pulse is used to prepare the system by exciting it from the ground state ($S_0$) to the valence excited state ($\pi\sigma^*$) of the molecule.

The interaction of the molecule with the pump-pulse is included in the initial Hamiltonian, and the system after the interaction is called the prepared system. The Hamiltonian of the corresponding system can be written as,
\begin{equation}
    \hat{H_I} = \hat{H}_0 + \hat{H}_{U}, \label{mod_hi}
\end{equation}
where $\hat{H}_0$ is the molecular Hamiltonian for the valence states, $\hat{H}_{U}$ is the Hamiltonian which considers the effects of pump-pulse on the molecule under the dipole approximation. The molecular Hamiltonian and the interaction with \ac{uv} pump-pulse reads:
\begin{align}
     \hat{H}_0 &= \begin{bmatrix} 
    \hat{T}+\hat{V}_0 & \hat{C}_{01}   \\
    \hat{C}_{01} & \hat{T} + \hat{V}_1 \\
\end{bmatrix} \label{mod_h0}\\
	\hat{H}_{U} &= - \mathrm{E_U} \cdot \cos(\omega_U t) \cdot f_U(t) \begin{bmatrix} 
    0 & \hat{\mu}_{01}   \\
    \hat{\mu}_{10} & 0 \\
\end{bmatrix} \label{mod_hu}
\end{align}
In the Hamiltonian matrix in Eq.\ \ref{mod_h0}, $\hat{T}$ and $\hat{V}$ represent the kinetic energy operator and \ac{pes} for a particular electronic state, respectively, and $\hat{C}_{01}$ represents the diabatic couplings between the states $S_0$ and $\pi\sigma^*$. The reduced mass $m_r=1809.5 \ \mathrm{a.u.}$ is used in the kinetic energy operator $T = (-1/2m_r)\nabla^2$, with $\nabla$ being the gradient operator with respect to the reaction coordinates $R_1$ and $R_2$. 
The $V$, $C$, and $\mu$ depend on the two reaction coordinates, $R_1$ and $R_2$. In the Hamiltonian describing the interaction with the pump, $\hat{\mu}_{01}$ and $\hat{\mu}_{10}$ represent the transition dipole moments between $S_0$ and $\pi\sigma^*$ states.
$\mathrm{E_U}$ represents the electric field strength of pump-pulse, $\omega_U$ is the carrier frequency of the pump-pulse and $f_U(t)= \exp(-t^2/2\sigma_U^2)$ is the Gaussian envelope of the laser pulse.

\section{Spectroscopic Signals} \label{sec:signal}

The expressions for the spectroscopic signals are discussed here, along with the modifications in Hamiltonian of the system. The electric field's effects from the pump pulse are calculated without any further approximation by including it in the system Hamiltonian (called prepared system) for all pump-probe signals.

\subsection{X-ray Absorption Signal}
In transient \ac{xas}, the absorption of an X-ray pulse, the interaction with a valence-to-core transition is recorded. The obtained spectrum contains information about the possible transitions.
We note that photoelectrons can also be generated in a competing process. However,
we neglect all the competing pathways and focus on the \ac{xas} signal.
Here we have included the pump pulse in the Hamiltonian which creates the 
initial state ($\hat H_I$ in Eq.\ \ref{mod_hi}). The interaction with the probe pulse is treated with time dependent
perturbation theory.
The total Hamiltonian can be written as follows,
\begin{equation}
    \hat{H}(t) = \hat{H}_I + \hat{H}_{int}(t)\,,  \label{xas_ht}
\end{equation}
where $\hat{H}_{int}(t)$ is the interaction Hamiltonian representing the interaction between probe-pulse and the prepared system. Under the \ac{rwa}, the time-dependent interaction Hamiltonian is
\begin{equation}
    \hat{H}_{int}(t) =  \hat{\mathcal{E}}_X(t)\hat{\mu}_X^{\dagger} + \hat{\mathcal{E}}_X^{\dagger}(t) \hat{\mu}_X  \label{xas_hint}
\end{equation}
with $\hat{\mu}_X\ (\hat{\mu}_X^{\dagger})$ being the transition dipole operator for describing the valence-to-core transition and $\hat{\mathcal{E}}_X(t)$ is the electric field operator
of the X-ray modes.
The field state is considered to be a coherent state initially. 
\begin{figure}[t]
  \centering
  \includegraphics[width=6cm]{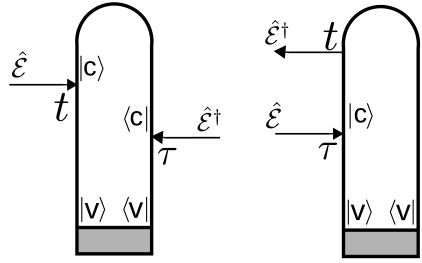}
  \caption{Loop diagrams corresponding to the \ac{xas} signal \cite{dorfman16rmp,mukamel10acmop}. The time runs along the loop from left bottom to right bottom. The interaction from left acts on the ket, and from the right acts on the bra. The arrows pointing inwards represent the absorption of the photon. The gray region defines the system prepared by the pump pulse, $\ket{\text{v}}$ represents the valence states, and $\ket{\text{c}}$ represents the core-hole states.} \label{fig:loop_xas}
\end{figure}
The signal for the \ac{xas} spectrum is defined as the integrated rate of change of the number of photons as a function of the absorption frequency $\omega_s$ and a pump-probe delay $T$, and can be expressed as follows,
\begin{equation}
    S (T,\omega_s) = \int_{-\infty}^{\infty} \mathrm{d}t \cfrac{\dd{\ev{N_s}}}{\dd{t}} \label{xas_sig}
\end{equation}
with $N_s$ being the photon number operator for the $s$th field-mode. The interaction Hamiltonian, $\hat{H}_{int}(t)$ in Eq.\ \ref{xas_ht}, appears in the signal expression as a consequence of the probe-pulse electric field $\hat{\mathcal{E}}_X$. 
The signal expression is then obtained by treating $\hat{H}_{int}(t)$ as the 
time-dependent perturbation.
The expression then reads  \cite{bennett16jctc,dorfman15pra,mukamel1999principles,kowalewski17cr},
\begin{align}\label{xas_sigt}
S(T, \omega_s) = &\cfrac{2}{\hbar^2} \mathcal{E}^*(\omega_s) 
  \Bigg [ \int_{-\infty}^{\infty} e^{i \omega_s (t-T)} \dd{t}  \\
& \times \int_{-\infty}^{\infty} \dd{\tau} \mathcal{E}_0 (\tau - T) e^{-i \omega_X (\tau-T)}  C_S(t,\tau) \Bigg]
  \nonumber
\end{align}
where $\omega_X$ is the carrier frequency of the X-ray probe-pulse, 
$\mathcal{E}_0(t) = \exp(-t^2/2\sigma_X^2)$ is its Gaussian envelope, and $\omega_s$ is the
frequency of the $s$th mode of the probe, and $C_S$ is the correlation function which contains the information about the interaction between the probe-pulse and the system, for time instances $\tau$ and $t$ as shown in Fig.\ \ref{fig:loop_xas}.
\begin{equation}
    C_S(t,\tau) = \bra{\psi_0}     \hat{\mu} (t)   \hat{\mu}^{\dagger} (\tau)  \ket{\psi_0} \label{xas_cf}
\end{equation}
 where $\ket{\psi_0}$ represents the superposition of the wave functions that evolve on valence states of the molecule according to the prepared-system Hamiltonian $\hat{H}_I$.
 Note that Eq. \ref{xas_sigt} describes both, absorption and stimulated emission according to the diagrams in Fig.\ \ref{fig:loop_xas}.

\subsection{X-ray Spontaneous Emission Signal}
\begin{figure}[t]
  \centering
  \includegraphics[width=3.1cm]{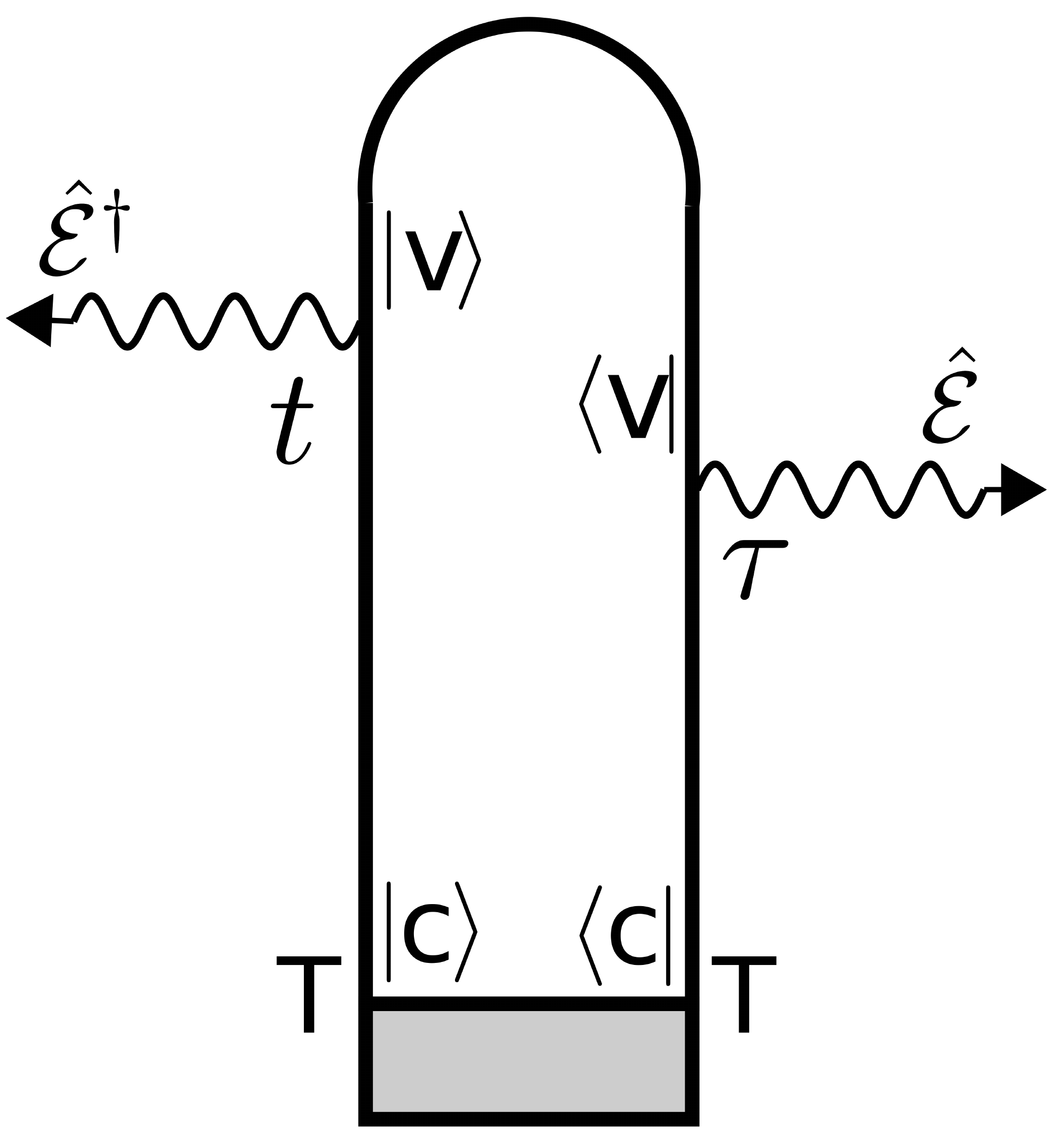}
  \caption{Loop diagram corresponding to the \ac{xses}. The gray region defines the interaction of system with pump-pulse as well as probe-pulse, and $\mathrm{T}$ is the time-delay between the pump-pulse and the probe-pulse.} \label{fig:loop_xses}
\end{figure} 
The probe pulse of a particular center frequency is used to excite the system from the valence states to the core-hole states. The $1s$ core-hole states of Nitrogen in pyrrole have a lifetime of $\approx$7\,fs \cite{inhester19jcp,hargenhahn01jpca,schoffler08sj} and hence spontaneous emission back
to the valence states occurs on this time scale. 
To keep the computational effort tractable and stay within Hilbert space, we introduce the core-hole lifetime as an empirical parameter. 
Here we also neglect the photoionization due to the X-ray probe pulse as well as the
subsequent Auger decay \cite{demekhin13jop,wolf20arxiv,bennett15fd} and instead solely focus on spontaneous emission process. Note that the Auger process can be a dominant decay channel for relaxation from core excited states in light elements \cite{haynes21np,wendin80ps}.
The total Hamiltonian for constructing the \ac{xses} signal is similar to the Hamiltonian in Eq.\ \ref{xas_ht}. The difference is that now both the pump and probe pulses are included in the initial Hamiltonian, and the subsequent spontaneous emission process is treated with perturbation theory. 
The molecular Hamiltonian, the pump interaction, and probe interaction are expressed as follows:
\begin{equation}
     \hat{H}_0 = \begin{bmatrix} 
    \hat{T}+\hat{V}_0 & \hat{C}_{01}  & 0 \\
    \hat{C}_{10} & \hat{T}+\hat{V}_1 &  0 \\
    0 & 0 &  \hat{T}+\hat{V}_2 \\
\end{bmatrix} \label{xses_h0}
\end{equation}
\begin{equation}
     \hat{H}_{U} = - \mathrm{E_U} \cdot \cos(\omega_U t) \cdot f_U(t) \begin{bmatrix} 
    0 & \hat{\mu}_{01}  & 0 \\
    \hat{\mu}_{10} & 0 &  0 \\
    0 & 0 &  0 \\
\end{bmatrix} \label{xses_hu}
\end{equation}
\begin{equation}
     \hat{H}_{X} = - \mathrm{E_X} \cdot \cos(\omega_X t) \cdot f_X(t) \begin{bmatrix} 
    0 & 0  & \hat{\mu}_{02} \\
    0 & 0 &  \hat{\mu}_{12} \\
    \hat{\mu}_{20} & \hat{\mu}_{21} &  0 \\
\end{bmatrix} \label{xses_hx}
\end{equation}
The notations in the molecular and pump field  Hamiltonians are identical to the notation used in Eq. \ref{mod_h0} and Eq. \ref{mod_hu} with $\hat{V}_2$ being a core-hole state. In the probe field Hamiltonian matrix $\hat{H}_{X}$, $\hat{\mu}$'s represents the dipole moment between the valence ($S_0$ and $\pi\sigma^*$) and core-hole ($C_1$) states, $\mathrm{E_X} =$ 2.06$\times \text{10}^{\text{11}}$\,V/m is the electric field strength of the probe pulse, $\omega_X$ is the carrier frequency of probe-pulse in X-ray regime, and $f_X(t)= \exp(-t^2/2\sigma_X^2)$ is the Gaussian envelope of the probe-pulse. The field operator of the spontaneously emitted photons takes the following form,
\begin{equation}
    \hat{\mathcal{E}}(t) = \sum_s \Bigg(   \cfrac{2\pi \hbar \omega_s}{\Omega}\Bigg)^{1/2} \hat{a}_s e^{ - i \omega_s t}    \label{xses_ex}
\end{equation}
where $\Omega$ is the quantization volume, and $\omega_s$ represents the frequency of the spontaneously emitted photon. The \ac{xses} signal is defined as the integrated rate of change of the number of photons as a function of the emission frequency $\omega_s$ and the pump-probe delay $T$. The signal expression then reads \cite{marx08pra},
\begin{equation}
\begin{aligned}
        S(T,\omega_s) =  -\cfrac{2}{\hbar\pi c^3} \real  \Bigg[ \int_{-\infty}^{\infty}& \dd{t} \int_0^{\infty} \omega_s^3 d\omega_s    \int_{-\infty}^t \dd{\tau} \\ 
        \times & e^{i \omega_s (t-\tau)}   e^{-\gamma t} C_E(\tau,t)     \Bigg] \label{xses_sigt}
\end{aligned}
\end{equation}
where $\gamma = 1/\delta t$, with $ \delta t =$ 7\,fs, is the decay rate for the core-hole states and $C_E$ is the correlation function corresponding to the diagram shown in Fig.\ \ref{fig:loop_xses}. 
The integral over the frequency $\omega_s$ is skipped in the final signal calculations as it is the variable for the frequency resolved detection. The two-time correlation function can be written as follows,
\begin{equation}
    C_E(\tau,t) = \bra{\psi_{0}(T)}  \hat{\mu}^{\dagger}(\tau)  \hat{\mu}(t) \ket{\psi_{0}(T)}\,, \label{xses_cf}
\end{equation}
where  $\ket{\psi_{0}(T)}$ represents the $C_1$ state which is populated at the pump-probe delay $T$ by X-ray pulse. 

\subsection{\ac{trc}}
\begin{figure}[t]
  \centering
  \includegraphics[width=3cm]{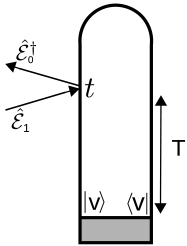}
  \caption{Loop diagram corresponding to the \ac{trc} signal. The gray region defines the system prepared by the pump pulse, $\ket{\text{v}}$ represents the valence states, and $\mathrm{T}$ defines the time-delay between pump-pulse and the hybrid probe-pulses.} \label{fig:loop_trc}
\end{figure} 
In contrast to \ac{xas} and \ac{xses}, \ac{trc} is a technique which is only sensitive to the vibronic coherences generated in the molecule.
It uses a hybrid, off-resonant probe-pulse sequence.
The signal consists of a Stokes and anti-Stokes type signal generated due to the redistribution of the photons between different modes.
The \ac{trc} spectrum is centered at the difference of the carrier frequencies of the hybrid probe pulses. The Raman shift indicates the energy difference between the involved valence states.
Here we assume that the probe pulses have the same carrier frequency. 
The \ac{trc} signal, which is constructed by perturbative treatment of the hybrid probe pulses, is represented by the following expression \cite{kowalewski15prl},
\begin{equation}
\begin{aligned}
        S(\omega_r,T) = 2 & \imaginary \int_{-\infty}^{\infty} \dd{t} e^{i\omega_r(t-T)} \mathcal{E}_0^*(\omega_r) \\ 
        & \times \mathcal{E}_1(t-T) \bra{\psi(t)} \hat{\alpha} \ket{\psi(t)} \label{trc_sigt}
\end{aligned}
\end{equation}
where $\psi(t)$ is the linear combination of the valence-states wave function at the pump-probe delay $T$, $\omega_r$ is the Raman frequency, $\mathcal{E}_0$ and $\mathcal{E}_1$ are the two probe pulses respectively, and $\hat{\alpha}$ represents the polarizability tensor of the molecule \cite{bernath20oup}:
\begin{equation}
    \hat{\alpha} = \begin{bmatrix} 
\alpha_{xx} & \alpha_{xy} & \alpha_{xz} \\
\alpha_{yx} & \alpha_{yy} & \alpha_{yz} \\
\alpha_{zx} & \alpha_{zy} & \alpha_{zz} \\
\end{bmatrix} \label{trc_alpha1}
\end{equation}
where each element $\alpha_{ij}$, with $i,j=x,y,z$ represents the directions of polarization.
Each tensor element is expanded in the basis of the valence states:
\begin{equation}
    \hat{\alpha}_{ij} = \begin{bmatrix} 
\alpha^{00} & \alpha^{01} \\
\alpha^{10} & \alpha^{11} \\
\end{bmatrix} \label{trc_alpha2}
\end{equation}
Every element $\alpha^{kn}$, with $k,n$ representing the initial and final valence states, depends on the nuclear coordinates $R_1$ and $R_2$, and the frequency of the probe-pulses $\omega$ \cite{bernath20oup}:
\begin{equation}
    \alpha_{ij}^{kn} = \cfrac{1}{\hbar} \sum_r \Bigg( \cfrac{\mu_{i,kr}\mu_{j,rn}}{\omega_{rn}-\omega} + \cfrac{\mu_{i,rn}\mu_{j,kr}}{\omega_{rk}+\omega} \Bigg)\label{trc_alpha3}
\end{equation}
where $\omega_{rn} = (E_r-E_n)/\hbar$ is the resonance frequency between the core-hole state $r$ and the valence state $n$.

The matrix from Eq.\ \ref{trc_alpha2} becomes non-Hermitian when Eq.\ \ref{trc_alpha3} is used. Here we make the
assumption that the energy difference between the valence states is small
compared to $\omega_{rn}-\omega$ and the off-diagonal elements can
be written as the average: $\alpha^{\prime} = (\alpha^{01}+\alpha^{10})/2$.
The new polarizability matrix, that is used in the calculations, then reads:
\begin{equation}
    \hat{\alpha}_{ij}^{\prime} = \begin{bmatrix} 
\alpha^{00} & \alpha^{\prime} \\
\alpha^{\prime} & \alpha^{11} \\
\end{bmatrix} \label{trc_alpha4} \,.
\end{equation}

\section{Methods} \label{sec:methods}
The ground state minimum structure of pyrrole, optimized at DFT/B3LYP/aug-cc-pVDZ level of theory, has been used as the reference geometry for computing the \acp{pes}. A rigid 2D-\ac{pes} scan has been performed by employing the complete self-consistent field (CASSCF) method along with the aug-cc-pVDZ basis set. Three states have been included in the state-averaged CASSCF for calculating the \acp{pes} corresponding to the electronic valance states. The active space is comprised of seven orbitals (5 $\pi$-orbitals of ring and a pair of $\sigma/\sigma^*$-orbitals of NH bond) and 8 electrons. To compute the core excited state energies involving the electronic transition from the 1s-orbital of nitrogen atom, complete active space configuration interaction (CASCI) method has been used. The prior-mentioned 3-state averaged CASSCF wave function, and the corresponding orbitals have been utilized to generate the configurations in the CASCI method. The diabatic \acp{pes} have been obtained using the quasi-diabatization procedure as implemented in MOLPRO-2019. The corresponding transformation matrix has been used to transform the dipole moments from adiabatic to diabatic basis. The program package MOLPRO-2019 \cite{MOLPRO-WIREs,MOLPRO_brief} has been used for the electronic structure calculations. 

The correlation functions in Eqs. \ref{xas_cf}, \ref{xses_cf}, and \ref{trc_sigt} have been calculated by a direct propagation simulation protocol \cite{Kowalewski15jcp}. 
The wave packet dynamics have been calculated with our in-house software QDng.
The Arnoldi method \cite{arnoldi51qam} was used to propagate the wave functions. 
The \acp{pes} and the numerical wave functions are represented on a grid with dimensions $256 \times 256$. 
A perfectly matched layer is used to absorb the wave packets at the boundary of the numerical grid.
The second derivative with respect to the reaction coordinates is calculated using the Fourier transform method. 
For all the propagations, a time step size of  50\,as has been used. 
Since the lifetime of nitrogen 1s core-hole states is $\approx$7\,fs,
the time-evolution of the core-hole states in the \ac{xses} signal is done until 12.2\,fs. 

\section{Results and Discussion} \label{sec:result}
\begin{figure}[t]
  \centering
  \includegraphics[width=8.5cm]{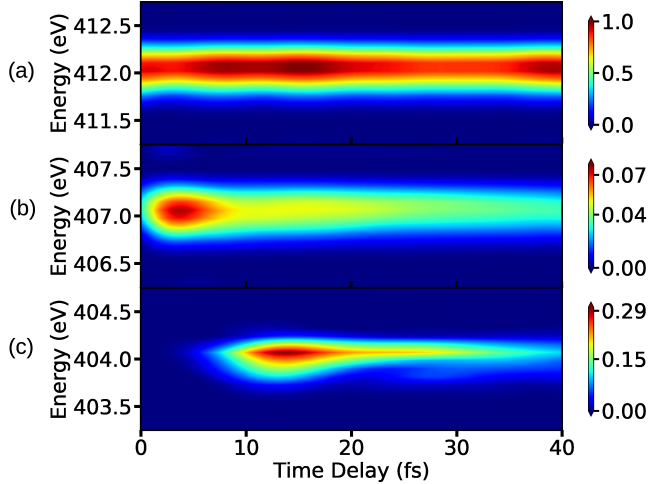}
  \caption{ The \ac{xas} spectra for the lowest core-hole for X-ray center frequencies with (a) $\omega_X =$ 412\,eV, (b) $\omega_X =$ 407\,eV, and (c) $\omega_X =$ 404\,eV.  The other \ac{uv} pulse and X-ray pulse parameters used for the signal calculations are as follows: $\mathrm{E_U} =$ 7.71$\times \text{10}^{\text{10}}$\,V/m , $\sigma_{U} =$ 1\,fs ($\text{FWHM} =$ 2.35\,fs), $\omega_{U} =$ 4.76\,eV, and $\sigma_{X} =$ 1.5\,fs ($\text{FWHM} =$ 3.53\,fs).} \label{fig:sig_xas1}
\end{figure}
The \ac{xas} signal is calculated by solving Eq.\ \ref{xas_sigt} for the core-hole states shown in Fig.\ \ref{fig:pes}. 
The $C_1$ state is energetically $3 \text{-} 4$\,eV apart 
from nearest core-hole state in the vicinity of the \ac{ci} and
thus the \ac{xas} signal is constructed only for the $C_1$ state.
This can be used to distinguish the various features in the spectrum.
The X-ray pulse has a finite width in the energy domain and thus the signal can be expected to show sharp features.
Using an appropriate pulse-width and center-frequency of the X-ray pulse, it is possible to probe absorption bands belonging to different energy regimes as shown in Fig.\ \ref{fig:pes}. 
The signals are shown in Fig.\ \ref{fig:sig_xas1} for three different X-Ray probe center frequencies,  corresponding to transitions A, B and C in Fig.\ \ref{fig:pes}. 
Note that, when all the core-hole states are considered for the signal construction, the transitions corresponding to A and B for $C_1$ state will be resonant with transitions from other core-hole states as well.
The \ac{xas} spectrum shown in Fig.\ \ref{fig:sig_xas1}(a) corresponds to the transition from state $S_0$ to state $C_1$ in the energy regime denoted by A in Fig.\ \ref{fig:pes}.
The signal has almost constant intensity for all measured time-delays, while small variations in intensity originate from the pump pulse and the unbalanced branching ratio near the \ac{ci}. We note that the signal shows similar features as the \acl{atas} spectrum recorded for the carbon K-edge in $C_2H_4^+$ to study the conical intersection between the $D_0$ and $D_1$ states \cite{zinchenko21sj}.
This signal has higher intensity compared to signals corresponding to the transitions B and C  in Fig.\ \ref{fig:pes}. The reason being the population residing at the local minima of $S_0$ state, as shown in Fig.\ \ref{fig:pes}, is higher than the population transferred to the $\pi\sigma^*$ state by the pump pulse.
Hence the intensity of signal in Fig.\ \ref{fig:sig_xas1}(a) is used as a reference for signals corresponding to other energy regions. The signal shown in Fig.\ \ref{fig:sig_xas1}(b) correspond to the transition B in Fig.\ \ref{fig:pes} and it has highest intensity when pump-pulse is about to vanish and the intensity decreases when the wave packets starts to move away from the \ac{fc} point. The steady decrease in the intensity after some time indicates the presence of a local minima in the $\pi\sigma^*$ state. A certain amount of population is trapped in that minima that continues to leak over time.
\begin{figure}[t]
  \centering
  \includegraphics[width=8.5cm]{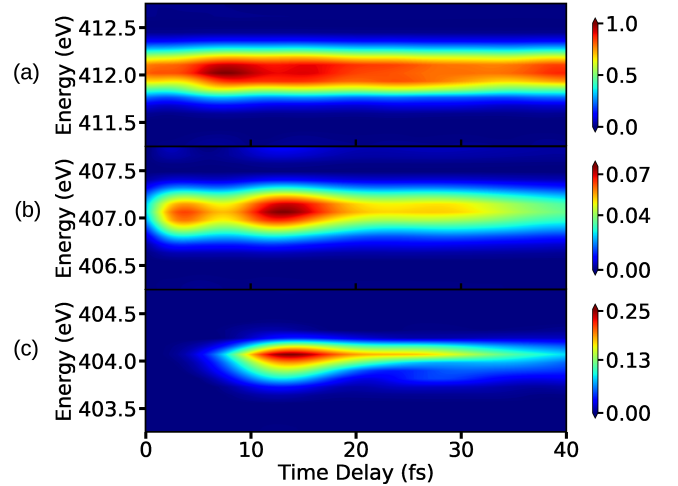}
  \caption{(a) The \ac{xas} spectra for five core-hole states for (a) $\omega_X =$ 412\,eV, (b) $\omega_X =$ 407\,eV and (c) $\omega_X =$ 404\,eV. The other pulse parameters for \ac{uv} pulse and X-ray pulse are same as the ones used in Fig.\ \ref{fig:sig_xas1}. The bump in (b) appears because the \ac{ci} region is resonant with $C_2$ for $\omega_X =$ 407\,eV, which was not present in Fig. \ref{fig:sig_xas1}(b).} 
\label{fig:sig_xas2}
\end{figure} 
Figure \ref{fig:sig_xas1}(c) indicates that the absorption is taking place near the \ac{ci}: the signals appears when the wave-packets on the $\pi\sigma^*$ state arrive in the vicinity of the \ac{ci} and it vanishes when the wave-packets move past the \ac{ci} region and move out of resonance with the probe-pulse.
The transition from $\pi\sigma^*$ state to $C_1$ near the \ac{ci} has a significantly higher intensity compared to transition between same states in the \ac{fc} region (transition B in Fig.\ \ref{fig:pes}), and it can be observed by comparing Fig.\ \ref{fig:sig_xas1}(b) and Fig.\ \ref{fig:sig_xas1}(c). 
The increase in the intensity of signal near the \ac{ci} indicates that the transition dipole-moment between the involved states increases in the vicinity of the \ac{ci}. 
The fact that there are parts of the wave packets retained in the local minimum at the \ac{fc} of the $\pi\sigma^*$ state supports the conclusion about increase in transition dipole-moment around \ac{ci}.
There is an increase in the intensity of the signal in Fig.\ \ref{fig:sig_xas1}(a), between 5 and 20\,fs and around 40\,fs, when the wave packets pass through the \ac{ci}. 
It is a result of the transfer of low energy wave-packets from $\pi \sigma^*$ to the $S_0$ state near \ac{ci}, that return to the minima of $S_0$ state. 
These vibrationally excited wave packets oscillate in the bounded region of the $S_0$ potential.
Note that it is not possible to observe the change in the nature of the $\sigma^*$ orbital along the dissociation coordinate in XAS spectrum. The X-ray probe frequency required to probe the transformation region is almost resonant with the absorption frequency of CI region. As a result the XAS signal from CI region dominates the XAS signal from the Rydberg-to-valence transformation region.

\begin{figure}[t]
  \centering
  \includegraphics[width=8.5cm]{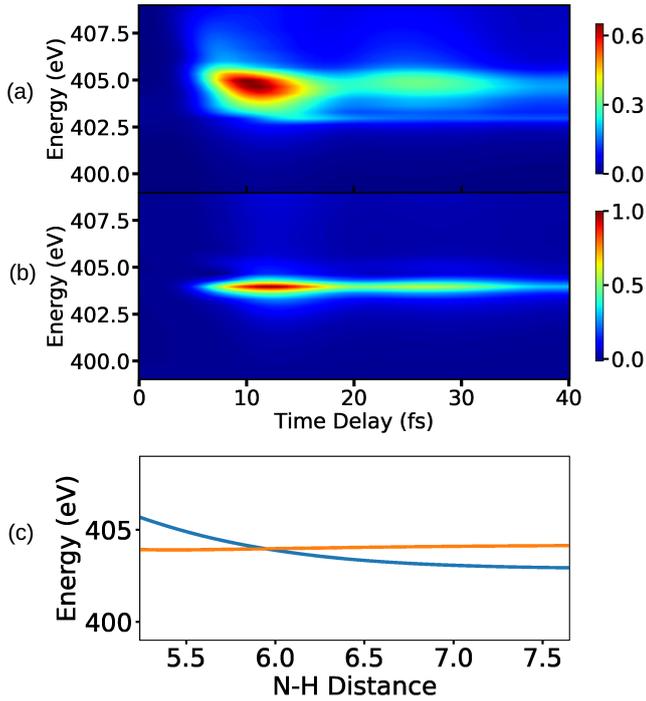}
  \caption{(a) The $S_0$ state contribution to the \ac{xses} spectrum for the $C_1$ state. (b) The $\pi\sigma^*$ state contribution to the \ac{xses} spectrum. (c) The separation between the $S_0$ and $C_1$ state (blue curve), and separation between $\pi\sigma^*$ and $C_1$ state (orange curve) as a function of one reaction coordinate for a fixed value of other reaction coordinate is shown here. The \ac{uv} pulse and X-ray pulse parameters used for the signal calculations are as follows: $\mathrm{E_U} =$ 7.71$\times \text{10}^{\text{10}}$\,V/m, $\sigma_{U} = $ 1\,fs ($\text{FWHM} = $ 2.35\,fs), $\omega_{U} = $ 4.76\,eV, and $\mathrm{E_X} =$ 2.06$\times \text{10}^{\text{11}}$\,V/m, $\sigma_{X} =$ 0.5\,fs ($\text{FWHM} =$ 1.18\,fs), $\omega_{X} =$ 403.3\,eV.} \label{fig:sig_xses1}
\end{figure}
\begin{figure}[t]
  \centering
  \includegraphics[width=8.5cm]{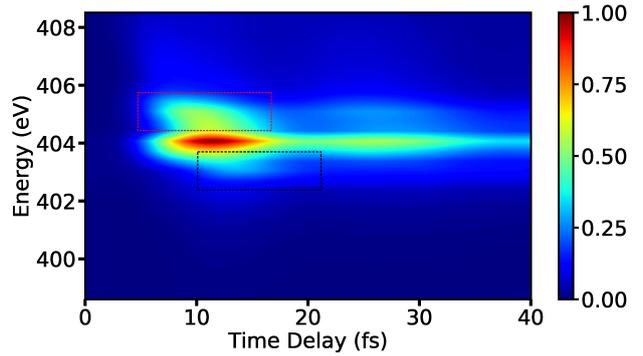}
  \caption{The total \ac{xses} spectrum for the $C_1$ state. The regions marked with red and black dotted boxes correspond to the $S_0$ state \ac{pes} before and after the \ac{ci}, respectively.} \label{fig:sig_xses2}
\end{figure}

The \ac{xas} spectra shown in Fig.\ \ref{fig:sig_xas2} contains the contributions from five core-hole states. 
The non-adiabatic couplings between the core-hole states, which have avoided crossings, are not included in the calculations. 
The avoided crossings are spatially well separated from the \ac{ci} between the valence states
and thus it is expected that these crossings do not significantly alter the spectrum.
When all five core-hole states (shown in Fig.\ \ref{fig:pes}) are used to construct the \ac{xas} signal, multiple valence-to-core transitions can be resonant with the same probe-pulse center-frequency. The transition C in Fig.\ \ref{fig:pes} has the lowest energy and it is the only transition resonant with the probe pulse with a center frequency 404\,eV. The number of resonant transitions increases by further increasing the carrier frequency. 

The \ac{xas} signal corresponding to the transitions taking place in the vicinity of \ac{ci} is shown in Fig.\ \ref{fig:sig_xas2}(c). Due to the resonance condition, signal in Fig.\ \ref{fig:sig_xas2}(c) has the
same shape as the signal for a single core-hole state has in Fig.\ \ref{fig:sig_xas1}(c).
The transition corresponding to the \ac{fc} region for the $\pi\sigma^*$ state is resonant with the transition between the $C_2$ state and the \ac{ci} region, and hence there is increase in intensity of the signal between $10 \ \mathrm{fs}$ and $20 \ \mathrm{fs}$ in Fig.\ \ref{fig:sig_xas2}(b). 
The overlap of multiple signals can also be seen in the Fig.\ \ref{fig:sig_xas2}(a) which correspond to the absorption from the $S_0$ state's local minimum. 
Since the intensity of the signal in Fig.\ \ref{fig:sig_xas2}(a) is used as a reference for other (Fig.\ \ref{fig:sig_xas2}(b) and (c)) signals, the spectrum in Fig.\ \ref{fig:sig_xas2}(c) has a lower intensity than the signal in Fig.\ \ref{fig:sig_xas1}(c). 
Apart from the change in peak intensity for absorption corresponding to 412\,eV in Fig.\ \ref{fig:sig_xas1} and \ref{fig:sig_xas2}, there is a noticable difference between the shape of the signals as well. 
The increase of intensity in the \ac{xas} spectrum in Fig.\ \ref{fig:sig_xas2}(c) at around 8\,fs, 
can be interpreted as an arrival of the wave packet in the \ac{ci} region.

The \ac{xses} spectrum can be used to complement the findings from the \ac{xas} signal and to 
observe the  energy gap between the valance states directly. 
The \ac{xses} spectrum is constructed by solving Eq.\ \ref{xses_sigt}. 
For the sake of clarity, the \ac{xses} signals are constructed for both $S_0$ and $\pi\sigma^*$ states separately. 
In the simulated spectra, the contribution to the total signal from each state can be evaluated. 

The spectrum shown in Fig.\ \ref{fig:sig_xses1}(a) corresponds to the signal of the $S_0$ final state and it gets stronger when the wave packets approach the \ac{ci}. The signal decays when the location of
the wave packets is not resonant anymore with the probe pulse.
This behavior follows a trend between 5\,fs and 20\,fs which corresponds to the blue curve in the Fig.\ \ref{fig:sig_xses1}(c) representing the separation between the $S_0$ state and the $C_1$ state as a function of a reaction coordinate. 
The signal that corresponds to the $\pi\sigma^*$ state is shown in Fig.\ \ref{fig:sig_xses1}(b) and the increase and decrease in intensity over the time shows a  similar behavior as in the case of the $S_0$ state. 

The change in intensity as a function of emission frequency and time delay in \ac{xses} spectra can be understood as follows: The spatial position of the wave packets on a \ac{pes} relates them to 
a possible transition frequency for a particular state. 
If the energy of an excitation pulse is resonant or close to resonance for that transition frequency, the excitations will be maximum which would result in higher emission. Since the energy of the X-ray pulse is 403.3\,eV, the signal will have the highest intensity when the wave packets reach the region before the \ac{ci} which has resonance frequency of around 404\,eV. For a specific delay, the strength of the signal would depend upon the transition dipole moment between the valence and core-hole states. 
Out of the two valence states, the $\pi\sigma^*$ state has higher amplitude of transition dipole-moment to the $C_1$ state and hence will have higher intensity of \ac{xses} signal.

As can be seen in Fig.\ \ref{fig:sig_xses1}(c), the separation between the $\pi\sigma^*$ state and the $C_1$ state is almost constant (represented by orange curve) and a similar feature is found in the \ac{xses} spectrum in Fig.\ \ref{fig:sig_xses1}(b). 
This indicates that the \ac{xses} may be able to project the relative shapes of the \acp{pes} of involved states. 
The total \ac{xses} spectrum which includes both the valence states is shown in the Fig.\ \ref{fig:sig_xses2}. The region shown in the red box corresponds to the $S_0$ state before the \ac{ci}, and the region shown in the black box corresponds to the same state but after the \ac{ci}. 
The signal in the black box starts to develop around 10\,fs, which matches the time when the system reaches the \ac{ci} as predicted by the \ac{xas} signal in Fig.\ \ref{fig:sig_xas1}. 
The valence states can be seen to approach each other before 10\,fs and move apart subsequently.

\begin{figure}[t]
  \centering
  \includegraphics[width=8cm]{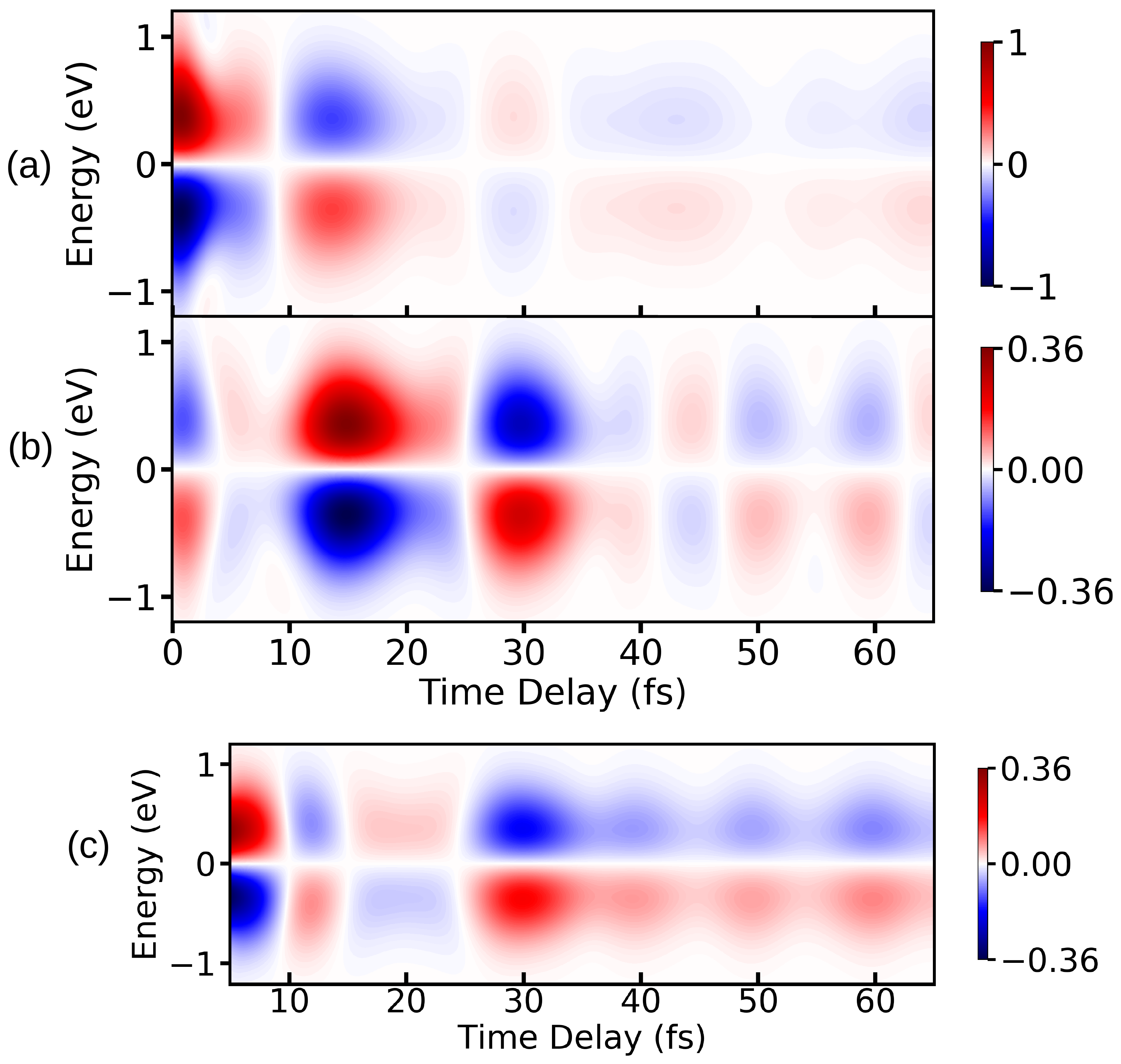}
  \caption{The \ac{trc} spectra for (a) $\alpha_{xx}$ and (b) $\alpha_{yy}$ (c) $\alpha_{xx}$ + $\alpha_{yy}$ are shown here. Note that the time-delay axis in (c) is cropped before 5\,fs to enhance the features of the signal. The intensity axes for all the figures are normalized with respect to the signal with highest intensity for the sake of comparison. The \ac{uv} pulse parameters are as follows: $\mathrm{E_U} =$ 2.06$\times \text{10}^{\text{10}}$\,V/m, $\sigma_{U} =$ 1.2\,fs, $\omega_{U} =$ 4.76\,eV. The parameters for the first probe-pulse $\mathcal{E}_0$ and second probe-pulse $\mathcal{E}_1$ are as follows: $\sigma_0 =$ 0.9\,fs ($\text{FWHM} =$ 2.12 \,fs), $\omega_{X} =$ 401\,eV, and $\sigma_1 =$ 1.7\,fs ($\text{FWHM} =$ 4\,fs).} \label{fig:sig_trc}
\end{figure}

Assuming that the transition dipole moments between the two valence states and the core-hole states are comparable to each other, the \ac{xses} signal can be used to detect the two \acp{pes} approaching each other before the \ac{ci} and move apart after the \ac{ci}. The only necessary condition is that the energy of the X-ray pulse should be resonant with the core-hole states in \ac{ci} region.

We now discuss the \ac{trc} spectrum. Figure \ref{fig:sig_trc} shows the \ac{trc} spectrum, which
has been calculated using Eqs.\ \ref{trc_sigt} and \ref{trc_alpha4}.
The \ac{trc} spectra for the $xx$ and $yy$ components of the X-ray polarizability tensor, 
$\alpha_{xx}$ and $\alpha_{yy}$ are shown separately in Figs.\ \ref{fig:sig_trc}(a) and (b) along with the sum of both components in Fig.\ \ref{fig:sig_trc}(c). Note that the sum spectrum corresponds to a
rotational averaged spectrum, which would be observed in a gas-phase experiment.
The spectrum corresponding to 
$\alpha_{zz}$ is not shown here because it is five orders of magnitude weaker.
The \ac{trc} spectrum without the \ac{uv} pump pulse has been subtracted from that of the prepared system with the \ac{uv} pump pulse to remove the contribution from ground state vibrational Raman transitions.

The peak before 5\,fs stems from the vibrational coherences created by the pump-pulse. 
At around 5-10\,fs the wave packet reaches the \ac{ci} region where the non-adiabatic couplings
come into effect. Inspection of Figs.\ \ref{fig:sig_trc}(a) and (b) shows an increase in intensity,
indicating a build up of coherence. 

Between 10 and 30\,fs the effects of the \ac{ci} can be seen in the spectrum. Both components in
Figs.\ \ref{fig:sig_trc}(a) and (b) show an increase in intensity. The $xx$ component in Fig.\ \ref{fig:sig_trc}(a) peaks between 10 and 20\,fs and falls off at later times.
The $yy$ component in Fig.\ \ref{fig:sig_trc}(b) also peaks between 10 and 20\,fs but has another
strong peak at $\approx$30\,fs.
This increase can
be explained as the build up in coherence as well as an increase in polarizability.
The $xx$ component of the polarizability of the $\pi\sigma^*$ peaks
before the \ac{ci} (at short NH bond distances) and goes to nearly zero after the wave packet
has passed through the \ac{ci}. The $yy$ component shows the opposite behavior: it is near zero before the wave packet reaches the \ac{ci} and peaks at larger NH bond distances. This explains the peak 
in Fig.\ \ref{fig:sig_trc}(b) at 30\,fs.
\begin{figure}[t]
  \centering
  \includegraphics[width=8cm]{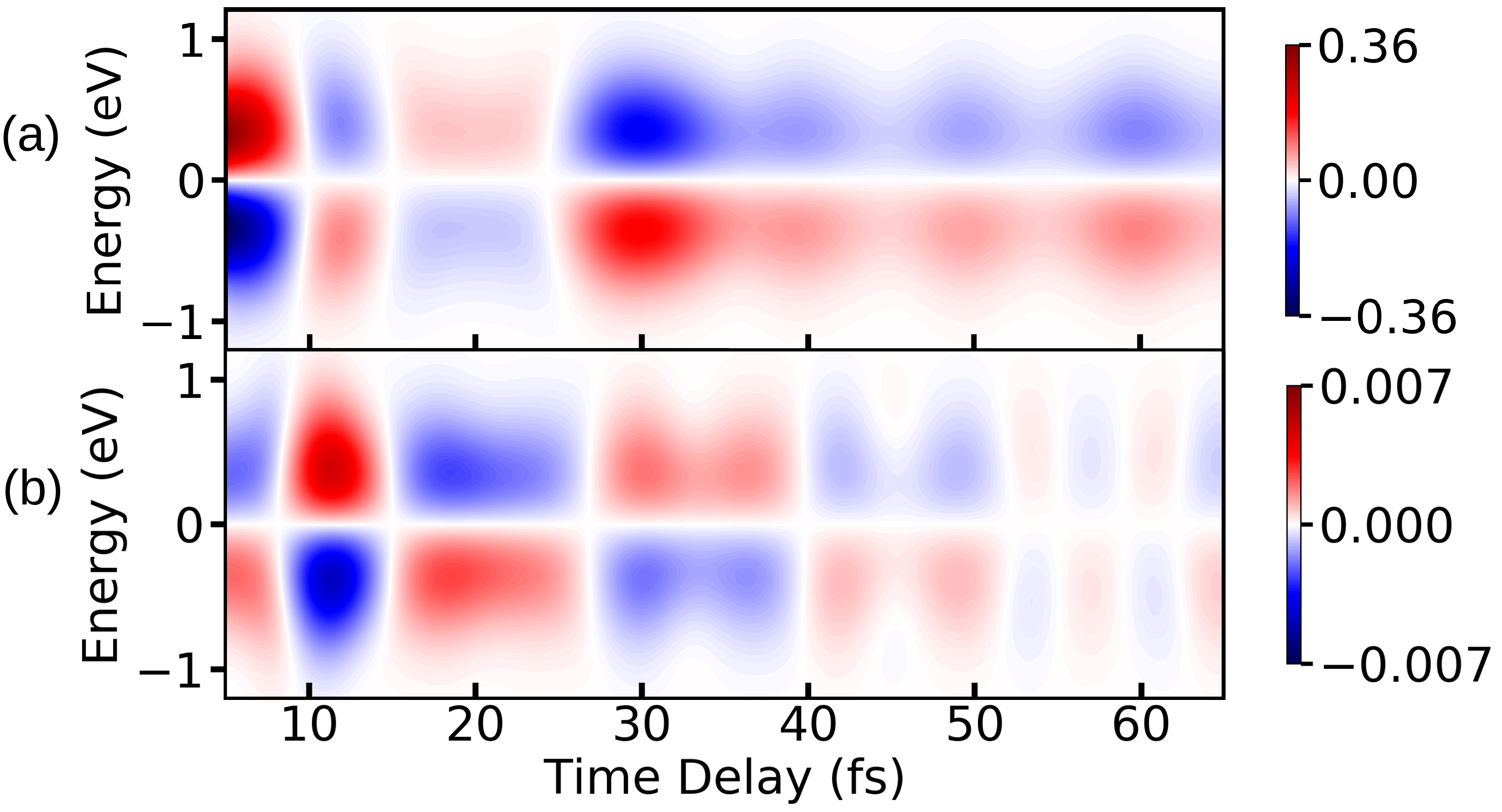}
  \caption{The TRUECARS signal for (a) vibrational coherence and (b) electronic coherence are shown. The intensity axes are normalized with respect to the TRUECARS signal in Fig.\ \ref{fig:sig_trc}(a).} \label{fig:sig_trc2}
\end{figure}
The Stokes and anti-Stokes patterns of the spectra 
for the $xx$ and $yy$ components are out of phase, leading to a partial cancellation of 
the signal in Fig.\ \ref{fig:sig_trc}(c). However, the oscillation patterns still have the same
period. The strength of the \ac{trc} signal depends on the magnitude of the matrix elements of the polarizability operator (see Eq.\ \ref{trc_alpha4}) along with the magnitude of the coherence generated. The diagonal elements of the polarizability operator constitute the vibrational coherence, while the off-diagonal elements constitute the electronic coherence in the spectrum. 
The \ac{trc} spectra are constructed separately for the vibrational coherence and the electronic coherence to compare the effect on the overall signal.
The \ac{trc} signal corresponding to electronic coherence is shown in Fig.\ \ref{fig:sig_trc2}(b), and it has a maximum intensity around 10\,fs which is the time-instance when the wave packets start reaching the \ac{ci}. The \ac{trc} signal corresponding to the vibrational coherence is shown in Fig.\ \ref{fig:sig_trc2}(a). Inspection of the polariziabilty matrix elements (see Eq.\ \ref{trc_alpha4}) also reveals that the diagonal elements are $\approx$2 times larger than the off-diagonal elements. Thus, the electronic coherences contribute significantly less than the vibrational coherences to the overall \ac{trc} signal.
Hence, the key signature of the \ac{ci}, i.e. electronic coherence generation, is not clearly visible in the \ac{trc} signal for pyrrole.
Note that the vibrational and electronic degrees of freedom are strongly mixed in the direct vicinity of the \ac{ci} and thus one can not clearly distinguish between vibrational and electronic coherences.

After 30\,fs the intensity of the signal decreases due to the dissociation. Note that this decay is mainly caused by absorption of the wave packet at the grid boundary (see Fig.\ \ref{fig:pops}).
In these particular examples of the \ac{trc} spectrum does not show the expected frequency resolution
and time evolution of the energy gap between the valence states can not be visualized.
This is mainly due to the choice of pulse parameters which are necessary to resolve the spectral features in the time domain.

\section{Conclusion} \label{sec:conclude}
We have studied the \ac{ci} in pyrrole using three different spectroscopy techniques, namely \ac{xas}, \ac{xses}, and \ac{trc}. 
A set of transient \ac{xas} spectra was constructed by selecting the dominant states involved in the non-adiabatic dynamics ($S_0$, $\pi\sigma^*$) and the transitions to the first nitrogen 1s core hole state. 
We could show that the signature of the \ac{ci} in the spectrum consists of a resonance and an increase in signal intensity:
For the lowest probe energy, the probe becomes resonant with the first core-hole state and a peak appears in the spectrum when the wave packet reaches the \ac{ci} region.
The transition dipole moment peaks in the vicinity of the \ac{ci} leading to an increased absorption indicating the time
it takes for the wave packet to reach the \ac{ci}, allowing to narrow down the timing further. Due to the unbalanced branching of the wave packet at the \ac{ci}, we see mainly the signature of the $\pi\sigma^*$ state in spectrum.

The time-resolved \ac{xses} spectrum was constructed using an X-ray probe-pulse, which has a frequency that is resonant with the valence-to-core transition corresponding to the \ac{ci} region. The $\pi\sigma^*$ state and the $C_1$ state have a similar shape and thus a constant energy difference in the vicinity of \ac{ci}.
Consequently, the spontaneous emission peaks at a rather constant photon energy. 
In contrast to the \ac{xas} spectrum,
the branching ratio at the \ac{ci} is not a decisive factor. 
The spontaneous decay from the nitrogen 1s core hole state has a similar
transition moment to $S_0$ and the $\pi\sigma^*$ state.
As a result, one can now see a signature of the curve  crossing in the \ac{xses} spectrum. 

For the third technique we have constructed the \ac{trc} signal. Here we explicitly probe
the vibronic coherences that are generated in the molecule by means of a linear off-resonant Raman process.
The \ac{trc} signal depends on the X-ray transition polarizability and the magnitude of the electronic and vibrational coherences of the excited system.
The simulated spectrum displays peaks  between 10\,fs and 20\,fs which hint at non-adiabatic dynamics
due to the \ac{ci}.
The spectra originating from the two polarizability tensor components are out of phase and the
overall signal is diminished.
The spectral features correlate with the signatures of the \ac{xas} and \ac{xses} spectra 
putting the passage through the \ac{ci} consistently between 10\,fs and 20\,fs.
We note that the \ac{trc} spectrum is dominated by the vibrational coherences and the increase of the polarizability near the \ac{ci} rather than the electronic coherences.

In conclusion, we have shown that the existence of a \ac{ci} in a molecule and its temporal appearance
can be identified by a combination of multiple methods. Each technique on its own may deliver a spectrum
that could potentially be difficult to interpret. The presented techniques probe different properties of
the molecule and thus their combination gives a more complete picture. We also note that the combination
is also potential strategy to bypass the high demand on the pulse bandwidth to map out the rapidly changing
energy near the \ac{ci}. The pulses used in our simulations were longer than 1\,fs thus explicitly avoiding attosecond pulses.

\section*{Acknowledgements} \label{sec:acknowledgements}
Support from the Swedish Research Council (Grant VR 2018-05346) is acknowledged.

\section*{Data Availability} \label{sec:dataavail}
The data is available on request from the authors.

\bibliography{report}   

\bibliographystyle{unsrt}

\end{document}